\documentclass[12pt]{article}

\usepackage{scicite}
\usepackage{amsmath}

\usepackage{times}

\title{Reducing cartel recruitment and upholding the rule of law is the only way to reduce violence in Mexico}

\author
{Raul Rojas$^{1\ast}$\\
\\
\normalsize{$^{1}$Department of Mathematics and Statistics, University of Nevada Reno,}\\
\normalsize{1664 N. Virginia Street, Reno, NV 89557, USA}\\
\\
\normalsize{$^\ast$E-mail:  rrojasgonzalez@unr.edu.}
}

\date{}

\begin{document} 


\baselineskip24pt


\maketitle


\section{Introduction}

Recently, the paper I want to comment on was published in Science\cite{cartels}. The main conclusion is also the title of the paper:  “Reducing cartel recruitment is the only way to lower violence in Mexico”. The authors examine the evolution of the total number of people recruited by  Mexican cartels and compare the current trend against two policy scenarios: (a) halving the recruitment rate of the cartels, or (b) doubling the rate at which cartel members are “incapacitated” (they go to prison). Of the two policies, the one that halves the recruitment rate wins because it leads to a reduction in the number of cartel members (the second policy reduces the growth rate of the cartels, but they still grow).

The curious reader will immediately ask: why not do both, a+b? Law enforcement is responsible for “incapacitation”, while social programs can try to keep young people out of gangs. The two policies are certainly complementary and
not exclusive. Therefore, the title and conclusion  of the paper are already biased since the simple alternative a+b exists. The title ("the only way") implicitly justifies the  \emph{laissez faire} policy of the current Mexican government, which is based on the false assumption that avoiding confrontation with the cartels would  somehow lead to a reduction in homicides. This has not
happened. The last five years have been the bloodiest in Mexico's modern history. In 2022, there were 32,223 homicides \cite{inegi} and many thousands of disappeared and missing persons.

Unfortunately, this is not the only thing wrong with this paper. The premise of the study is really simple: from the statistics of homicides in Mexico, one can postulate that a certain percentage of deaths correspond to cartel members. Also, the number of incarcerated narcos can be estimated from official or news reports. Since the sum of deaths plus incapacitated cartel members is proportional to the total number of members, if we postulate the coefficient of proportionality, we can then roughly calculate the total manpower of the cartels. The authors estimate that the cartels employed about 175,000 people in 2022, which would make them the fifth largest employer in the country (although they are not a coherent corporation).\footnote{This  ignores the fact that PEMEX pay full salaries to 140,000 active and 110,000 “inactive” employees. But this is just a detail, and I mention it only because the paper uses a bar chart  to rank the cartels among Mexican companies. The paper also ignores  public sector employment. There are more than half a million teachers in Mexico. There were  60 million people employed in Mexico in 2023 (both in the formal and in  the informal sector), so  a more 
useful statistic would be to say that for every 1000 workers, the cartels employ three people, that would be  1.3 persons per 1000 inhabitants. To give some perspective, let us mention that the US has an incarceration rate of 0.53 persons per 1000 inhabitants.}

\section{Technical criticism}

The main problem with this paper are the assumptions and the number of constants that need to be estimated to fit the distribution of cartel sizes. In the paper, the cartel size $C_i$ is the total number of people
working for the $i$-th cartel. The sum of the people employed by all cartels is $C$. The main equation in \cite{cartels} is Eq. 1, which is also the only one in the paper (the supplementary material does not add much more to it).

According to Eq. 1 the derivative $\dot{C}_i$ of cartel size $C_i$ changes according to
\begin{equation}
\dot{C}_i= \underbrace{\rho C_i}_{\text{recruitment}}- \underbrace{\eta \frac{C_i}{C}}_{\text{incapacitation}} - \underbrace{\theta \sum_{j\neq i}^{k} C_iC_jS_{ij} }_{\text{conflict}} - \underbrace{\omega C_i^2}_{\text{saturation}} \label{eq1}
\end{equation}
where  $\rho, \eta, \theta, \omega$ are coefficients to be estimated, and there are $k=150$ cartels. Recruitment $\rho C_i$ is directly proportional to the size of a cartel, which cannot grow indefinitely due to $\omega$. Incapacitation is proportional to the percentage of the size of a cartel relative to the total number of cartel members ($C$), and conflict is the number of people from cartel $i$ that the other
cartels kill.
Let us first comment  on the paper's assumptions regarding this equation:

\begin{itemize}
\item The authors include 150 cartels, a surprising number for a model. One would think that including only the most important cartels would be
simpler,  require fewer constants, and  produce a more robust result.

\item The positive or zero constants $S_{ij}$ represent the strength of conflict between cartel $i$ and cartel $j$. Thus, the model contains $150^2 - 150$ conflict parameters. That is   22350  parameters that we need to assign a value to.

\item Since this is an extraordinary number of parameters the authors make the arbitrary assumption that $S_{ij}$ is equal to $(R_{ij}+\varepsilon)/(A_{i}+1)$, where $R_{ij}$ represents the number of states
where cartels $i$ and $j$ are in conflict and $A_{i}$ is the number of alliances that cartel $i$ has formed. This is called its ``strength''. Of course, this is an arbitrary assumption, and the data sources are very 
fragmentary and questionably. Probably not even the Drug Enforcement Administration can provide such
numbers. Also, the simulation period is ten years (2012 to 2022) and we know well that conflicts and alliances between these criminal groups are fluid and highly volatile. These $S_{ij}$ are not really constants,
and they  never will be.

\item To simulate the growth of the cartels, using the proposed differential equations, the size of the 150 cartels is seeded for $t=0$ using a power law, which is just $C_i = D/i$, where $D$ is the initial number of  members of all cartels $C(0)$ (year zero is 2012), divided by the sum of weights $1+1/2+1/3+ \ldots 1/150$. The assumption is that cartel 2 is half the size of cartel 1, cartel three is one third the size, and so on. Of course, this is another arbitrary assumption since a power law can have an exponent other than  $-1$.   Table 2 in the Supplementary Material shows the results of the simulation up to 2022.  The final estimated size of the cartels does not follow the  power law postulated for $t=0$. After the first 13 cartels, ranked by size, the remaining cartels are hardly distinguishable in  size.  The last 130 cartels  now have 55,000 members instead of the 40,000 they should have according to the assumed power law distribution.   If we do not find the postulated power law in the simulation for 2022, why should it be present in 2012?

\item There is another problem with the seeding of the names of the cartels in the power law distribution. Since the values of the constants $A_i$ represent no more than a very weak guess, many possible permutations of cartel names (or their ``strengths'') could be used. The order of the cartel names in the results cannot be taken seriously, after  the six or seven  main cartels have been listed. The more powerful cartels listed in the ``2020 National Drug Threat Assessment''\cite{threat} do not correspond to the order of cartels in Table 2 . One must assume that the Drug Enforcement Administration produces its reports based on the best available intelligence.
Also, with the assumed  power law distribution, the cartels 100 to 150 have from 280 to 190 members, which qualifies them as criminal gangs, but hardly as cartels capable of holding a significant territory.

\item The constants $\rho, \eta, \theta, \omega$ are fitted by setting two other constants, $f$ and $g$. The constant $f$ represents the percentage of cartel deaths out of all violent deaths in
Mexico, during a year. The constant $g$ represents the percentage of cartel-related ``incapacitations'' among all incapacitations. These two numbers are probably impossible to obtain from the
official statistics, so the authors just guess that $f=0.1$ and $g=0.05$. The only justification the authors offer for this choice (in the Supplementary Material) is that they varied slightly both numbers and ``the variation does not alter the results significantly''. This is wrong, because the explicit argument is that the  results obtained justify the results.

\item Given $f$ and $g$, the authors then compare the number of cartel casualties from the model with the estimated number of casualties from data, and the number of incapacitations from the model with the estimated number of incapacitations (based on the annual data and the values selected for $f$ and $g$). They minimize the sum of the squared differences and from this they say that they obtain the values of the parameters $\rho, \eta, \theta, \omega$.

\item It is obvious that any model based on 22350 constants, whose numerical value is determined by a seemingly arbitrary criterion, and where four variables have to be optimized, based on 150 
differential equations, and where the error term depends \emph{crucially} on values for $f$ and $g$ set by hand, will produce wildly different results if just a few assumptions are changed. 

\item To justify and visualize the model, the paper considers a country with one or two cartels (in the Supplementary Material). This is just an empty exercise.

\end{itemize}

\section{One model for $C$, a simulation for each $C_i$}

In fact, if we look more closely at Eq. 1, we see that it represents a model for the evolution of the total labor $C$ employed by the cartels, and the
simulation of the 150 differential equations simply allocates that total $C$ among the 150 different cartels, producing $C_i$ for each $i$.
Stripping the model down 
to its bare bones shows us that the model is actually underdetermined.

Eq. \ref{eq1} can be interpreted as a difference equation. Adding $\Delta {C}_i$ for $i=i$ to $k$, and ignoring the quadratic term (setting $\omega=0$)\footnote{For a cartel with 28000 members the square of $C_i$ is 784 million, so that $\omega$ has to be very small in order not to make the growth rate negative. As a global parameter, $\omega$ is superfluous because it \emph{must} be extremely small.}, we
get
\begin{equation}
 \Delta {C} = \rho C - \eta -\theta \sum_{ i}^{k} \sum_{j\neq i}^{k} C_iC_jS_{ij} , 
 \end {equation}
  The variable $\eta$ is the total number of incapacitations in the period under consideration.
 The authors use $\theta$ to force the number of deaths they 
compute in their simulation to be equal to the actual number of deaths. Setting $\theta = V/\sum_{ i}^{k} \sum_{j\neq i}^{k} C_iC_jS_{ij} $, where $V$ is the total number of
 cartel members killed in the period, we have the desired effect, and we obtain the equation
\begin{equation}
 \Delta {C} = \rho C - \eta -V , 
 \end {equation}

 If we separate the computation of individual $C_i$'s and stick to computing the total size of the cartels $C$,
we do not need $\theta$.
To keep the total membership of the cartels stable over time, all that is 
 needed  is $\rho C = \eta +V$. This is an obvious statement. The cartels need to recruit at least as many people as they loose each year, just to keep their combined
size stable.

Notice, however, that we do not know the value of $\rho$. We might have $\rho C = (0.01)(100,000)= (0.02)(50,000)$. Given the assumptions, we cannot choose between
$C=100,000$ or $C=50,000$. That is, given the observables $\eta$ and $V$, we can only estimate $\rho C$, but not $C$ itself. Even with no growth at all,
the cartels could  lose more people each year if they could recruit enough replacements.

The authors estimate that 230 persons left the cartels every week in 2021, that is, 12,000 per year. Since they estimate that there are 175,000 cartel members, this would
represent a probability of being captured or killed of 6.8\% per year. The recruitment rate must be higher than this percentage just to keep the number of cartel members stable.

So imagine a scenario where the original $C$ does not increase from year to year, but $\rho$ does. Any increase in the annual number of deaths and incapacitations
could be absorbed by an increase in $\rho$. The fluctuation in C (given by $\rho C$) would be offset by an increased $\eta$ and $V$. You can have more annual deaths and
incapacitations without an increase in $C$.

The most likely scenario is that both $C$ and $\rho$ have increased in the last ten years, but this model has no way of distinguishing between the two
effects without additional assumptions. Also, there is no real justification in the paper for the initial total cartel size in 2012 of 115,000 persons. At the initial  time $t=2012$  the simulation
cannot be used, and therefore this is just a guess in order to start the simulation. But if the authors can guess $C(2012)$, this shows they could have
derived $C(2022)$ from their  assumptions  about $f$ and $g$ and the time series of  homicide and incarceration rates, without any simulation. And if they allocated $C(2012)$ among the cartels using
a power law, they could have done the same for $C(2022)$. The simulation of the ten years between 2012 and 2022 seems completely superfluous, given the arbitrary
assumptions about the constants $f, g, S_{ij}$.

What this section then shows is that Eq. 3, which determines the dynamics of $C$, does not depend on the distribution of cartel sizes. And the distribution of cartel
sizes cannot be obtained from Eq. 3 or Eq. 1. It is an artifact of the simulation, which has been seeded with arbitrary values for $S_{ij}$ and the initial power-law distribution for the $C_i$.

\section{Conclusions}

The main idea of  \cite{cartels} is very simple: the more cartel members that fall dead or are incarcerated each week, the larger the total number of cartel members.
Assuming a certain percentage of cartel members fall dead or are incapacitated each week, we can infer the total manpower employed by the cartels. The  power law
distribution takes care of seeding the cartels with a certain assumed size after they have been ranked. 

Because there are so many arbitrary assumptions in the model, and too
many parameters, an educated guess (``$x$ percent of the narcos die or go to jail every week'') would have produced an estimate from data \emph{as valid}
as the one produced by this  overparameterized model.

Of course, the conclusion that ``the only way'' to stop the cartels is to halve recruitment does not hold, because in the model itself, halving recruitment and doubling
incapacitacion is a  better alternative. Just look at Eq. 3. There are two policies that can reduce $\Delta C$, through $\rho$ or $\eta$, and they are not exclusive.

The main political problem in Mexico is that it is not a country with a ``rule of law'' tradition. The country is ranked  115th in the \emph{Rule of Law Index }of the World Justice
Project. The current policy of letting the cartels operate as long as they do not attack the military or do not cause too many civilian casualties has become
illusory and unsustainable, as the violence statistics show. The only way to reduce violence in Mexico is, first, to reduce recruitment, but second and more important, to uphold the rule
of law at all levels  ending the implicit tolerance of cartel activity in Mexico.

Moreover, great caution and restraint must be exercised in attempting to reduce very complex social problems to differential equations. 
The complexity of society is not  easily quantifiable.

\end{document}